\documentclass[useAMS,usenatbib]{mn2e}
\usepackage{graphicx}

\title[Mirror reflection vs. diffusion anisotropy]
{Effects of mirror reflection versus diffusion anisotropy  
on particle acceleration in oblique shocks}

\author[Y. S. Honda and M. Honda]{Y. S. Honda$^{1}$\thanks{E-mail:
yasuko@ktc.ac.jp (YSH)} and M. Honda$^{2}$\\
$^{1}$Kinki University Technical College, Kumano, Mie, 519-4395, Japan\\
$^{2}$Plasma Astrophysics Laboratory, Institute for Global Science, 
Mie 519-5203, Japan}
\begin{document}

\date{Accepted 2005 Received 2005 3 June; 
in original form 2005 January}

\pagerange{\pageref{firstpage}--\pageref{lastpage}} \pubyear{2005}

\maketitle

\label{firstpage}

\begin{abstract}
Cosmic ray particles are more rapidly accelerated in oblique shocks, 
with the magnetic field inclined with respect to the shock normal 
direction, than in parallel shocks, as a result of mirror reflection 
at the shock surface and slower diffusion in the shock normal direction. 
We investigate quantitatively how these effects contribute to reducing 
the acceleration time over the whole range of magnetic field 
inclinations. It is shown that, for quasi-perpendicular inclination, 
the mirror effect plays a remarkable role in reducing the 
acceleration time; whereas, at relatively small inclination, the 
anisotropic diffusion effect is dominant in reducing that time. These 
results are important for a detailed understanding of the 
mechanism of particle acceleration by an oblique shock in space and 
heliospheric plasmas.
\end{abstract}

\begin{keywords}
acceleration of particles --- diffusion --- magnetic fields 
--- shock waves --- cosmic rays.  
\end{keywords}

\section{Introduction}
Diffusive shock acceleration (DSA) is one of the most favourable 
mechanisms to reproduce the energy spectrum of observed cosmic rays 
(for a review, see Drury 1983). Shock waves accompanied by the 
magnetic fields with small fluctuations (Alfv\'en waves) are well 
established in space plasmas and operate as a powerful accelerator 
of charged particles \citep{blandford87}. In earlier works, 
substantial efforts were devoted to studies of acceleration by 
the simple `parallel shocks' in which both the magnetic field 
and the direction of plasma flow are parallel to the direction of 
shock propagation \citep{axford78,bell78,blandford78}. At the parallel 
shock fronts, all upstream particles are advected into the downstream 
region, gyrating around the mean magnetic field. According to the 
conventional resonant scattering theory, a particle with its gyroradius 
comparable to the turbulent Alfv\'en wavelength is resonantly scattered 
and migrates back upstream after many times of small-angle scattering.  
Cosmic ray particles acquire their energies in the process of transmission 
back and forth across the shock. Since the particles gain only a 
small amount of kinetic energy in each traversal of the shock, the 
acceleration efficiency depends largely on the rate of shock crossing 
of the particles.   

Relating to particle acceleration in the solar wind and the Earth's 
bow shock, the more general `oblique shocks', across magnetic field 
lines, have been studied \citep{toptyghin80, decker83}.  
Some researchers have argued that the acceleration 
efficiency is enhanced in oblique shocks, compared 
with that in parallel shocks \citep{jokipii87, ostrowski88}. 
If the particle momenta along the magnetic field, $p_{\|}$, are  
larger than a critical value, the particles gain energy solely  
via successive transmissions through the shock front, just as in the case 
of parallel shocks. However, when the value of $p_{\|}$ is smaller than 
the critical value, upstream particles cannot penetrate into the 
downstream region with stronger magnetic field and are reflected back 
into the upstream region with weaker field, having their pitch 
angles reversed. On the basis of the conservation of magnetic moment, the 
turnover point of the pitch angle 
is determined by the ratio of the upstream/downstream magnetic field 
strength. This `mirror reflection' results in significant reduction of  
acceleration time. Quite efficient acceleration is expected,  
in particular, for quasi-perpendicular shocks with larger 
inclination of magnetic field lines. 

It was also pointed out that the acceleration time for oblique 
shocks could be reduced owing to the anisotropy of the particle 
diffusion coefficient \citep{jokipii87}. 
The effective diffusion coefficient involved in the DSA time-scale 
can be represented by the tensor component normal to the shock 
surface, $\kappa_{{\rm n}}$, which is decomposed into components 
parallel ($\kappa_{\|}$) and perpendicular ($\kappa_{\bot}$) to the 
magnetic field (see also Section 2.1). In the special case of  
parallel shocks, the effective diffusion coefficient reduces to  
$\kappa_{{\rm n}}=\kappa_{\|}$. In the ordinary case of 
$\kappa_{\|}\gg\kappa_{\bot}$, reflecting one-dimensional (1D) 
properties in the conventional magnetohydrodynamic (MHD) description 
\citep{ostrowski88}, the value of $\kappa_{\rm n}$ decreases as the 
magnetic field inclination increases, and in perpendicular shocks, 
$\kappa_{\rm n}=\kappa_{\bot}$.  
Within the DSA framework, the smaller value of $\kappa_{\rm n}$ 
leads to a shorter acceleration time, and thereby a higher acceleration 
efficiency. 

Summarizing the above discussions, in oblique shocks there exist 
two possible effects contributing to the reduction of acceleration 
time: mirror reflection at the shock surface, and 
diffusion anisotropy. In previous works, these 
contributions were comprehensively treated with the `oblique effect'.  
In the present paper, we investigate the contribution of these effects 
separately and reveal which of these two effects contributes 
more dominantly to the acceleration time over the 
whole range of field inclination.  
For this purpose, we derive the expression for the acceleration time 
with and without the effects of mirror reflection.
Presuming that no particles are reflected at the oblique shock, 
we estimate the acceleration time {\it without mirror effects}, though 
still including the effects of diffusion anisotropy. 
To our knowledge, so far there has been no publication that quantitatively 
reveals the effects of anisotropy on the reduction of acceleration 
time for oblique shocks. 
Here we demonstrate that mirror reflection makes a 
dominant contribution to the rapid acceleration for highly inclined 
magnetic fields, whereas anisotropic diffusion becomes effective at 
smaller inclination angles. 

\section[]{Time-scales of oblique shock acceleration}
In the following description, we use a unit system in which the speed 
of light is $c=1$. Upstream and downstream quantities are denoted by 
the indices i=1 and 2, respectively. The subscripts $\|$ and 
$\bot$ refer to the mean magnetic field, and subscript n refers to 
the shock normal. 
The subscripts $\sigma$ indicate the processes of particle interaction 
with the shock: $\sigma=$r for reflection and $\sigma=$12 (21) 
for transmission of the particles from region 1 (2) to region 2 (1).  
For non-relativistic flow velocities, calculations related to 
particle energy gain are performed up to first order in $V_{i}$, 
where $V_{i}=U_{i}/\cos\theta_{i}$, and $U_{i}$ and $\theta_{i}$ are 
the flow velocity in the shock rest frame 
and the magnetic field inclination to the shock normal, respectively. 

\subsection{Effective diffusion coefficient}
To evaluate the DSA time-scale of cosmic rays, we need their spatial 
diffusion coefficient. According to the tensor transformation, 
the effective diffusion coefficient for the shock normal direction 
can be expressed as \citep{jokipii87} 
\begin{equation}
\kappa_{{\rm n}i}=\kappa_{\| i}\cos^{2}\theta_{i}
+\kappa_{\bot i}\sin^{2}\theta_{i}.
\label{eqn:kappa}
\end{equation}
Without going into the details of the scattering process, here we 
make use of the empirical scalings \citep{ostrowski88} 
\begin{equation}
\kappa_{\|}=\kappa_{\rm B}x ~~~~{\rm and}~~~~  \kappa_{\bot}=\kappa_{\rm B}/x.
\end{equation}
Here, the scaling factor $x(\geq 1)$ reflects the energy density of 
the mean to fluctuating magnetic fields, $(B/\delta B)^{2}$, and 
$\kappa_{\rm B}=r_{\rm g}v/3$ is the Bohm diffusion coefficient,  
where $r_{\rm g}$ and $v$ are the gyroradius and speed
of a test particle, respectively.  

Making the assumption that the magnetic moment of the particle is 
approximately conserved during the interaction with the discontinuous 
magnetic field at the shock front, we consider the situation in which 
small fluctuations are superimposed on the mean regular field, i.e. 
$x\gg 1$, in the vicinity of the shock (free crossing limit). On the 
other hand, in the case of $x\sim 1$ (diffusive limit), the fluctuations 
significantly affect the coherent gyromotion of the particle, and  
precise numerical simulations are required \citep{decker86}.     
For $x\gg 1$, we have the relation of 
$\kappa_{\bot}(=\kappa_{\|}/x^{2})\ll \kappa_{\|}$. 
This means that, for large angles $\theta_{i}>\tan^{-1}x$, 
the term involving $\kappa_{\bot}$ becomes dominant on the 
right-hand side of equation (1).                               
Thus, one finds that larger values of $x$ and $\theta_{i}$ 
are likely to lead to a shorter acceleration time. 
Here it is noted that the acceleration time cannot be infinitesimally 
small, because of the limits of $\kappa_{\rm n}$ and $\theta_{i}$ 
(see Section 2.2). 

The parallel/perpendicular diffusion of cosmic rays in turbulent 
plasmas is a long-standing issue and is not yet completely resolved.   
Analytic models are consistent with neither observations 
\citep{mazur00,zhang03} nor numerical simulations 
\citep{giacalone99,mace00} over a wide energy range.
This situation has changed since the development of non-linear 
guiding centre theory (NLGC) by \citet{matthaeus03}. Recent calculation 
of the NLGC diffusion coefficient suggests reasonable values of 
$x^{2}\ga 10^{3}$ for the typical solar wind parameters \citep{zank04}, 
which can be accomodated with the free crossing limit ($x\gg 1$). In 
many astrophysical 
environments, the perpendicular diffusion of cosmic rays still remains 
uncertain and is commonly inferred from fluctuations of the turbulent 
magnetic field. We will discuss this point in the conclusions.    

As long as the values of $x$ can be regarded as constants, the present 
consequences concerning the acceleration time are generic, apparently 
independent of the type of turbulence. However, for an 
application, when evaluating the maximum possible energy of a particle 
by equating the acceleration time with the competitive cooling time-scales, 
one should pay attention to the fact that, in general, $x$ depends on 
the ratio of $r_{\rm g}$ to the correlation length of turbulence, 
deviating from the simplistic scaling 
$x\sim (B/\delta B)^{2}$ \citep{honda04}. Here, note 
that, in the Bohm limit with $\nu=1$, the above dependence is found 
to be weak, so as to appear as logarithmic. For the present purpose, 
below, we assume $x$ to be a constant much larger than unity, reflecting 
weak turbulence, and thereby larger anisotropy of diffusion. 
 
\subsection{Particle acceleration including mirror effects}
Following the kinematic approach developed by \citet{ostrowski88}, 
for instruction we outline the derivation of the acceleration 
time for an oblique shock including the effects of both mirror 
reflection and diffusion anisotropy. For convenience, the 
energy and momentum of a test particle are transformed from the 
upstream/downstream rest frames to the de Hoffmann-Teller (HT) frame 
\citep{hoffmann50,hudson65}. Since the particle kinetic energy 
is invariant in the HT frame, where electric fields vanish,  
one can easily estimate the energy gain of the particle during 
the interaction with the shock, presuming the conservation of 
magnetic moment of the particle. Finally, all variables are 
transformed back to the original rest frames. 

In the HT frame, if the cosine of particle pitch angle, $\mu$, 
is in the range $0<\mu<\mu_{0}$, the upstream particles are 
reflected at the shock surface. Here, 
\[
\mu_{0}=(1-B_{1}/B_{2})^{1/2} 
\]
gives the critical value, where $B_{1}<B_{2}$ for the fast-mode 
oblique shock. In this case we have the following ratio of particle 
kinetic energies: 
\[
\frac{E_{r}}{E}=\gamma_{1}^{2}(1+V_{1}^{2}+2V_{1}\mu), \nonumber 
\]
where $E$ and $E_{r}$ are the particle energies (in the 
region 1 rest frame) before and after reflection, 
respectively, and $\gamma_{1}=(1-V_{1}^{2})^{-1/2}$. 
If $\mu_{0}<\mu\leq 1$, the particles are transmitted to region 2.
In this case, we have 
\[
\frac{E_{12}}{E}=\gamma_{1}\gamma_{2}\left\{1+V_{1}\mu-V_{2}\left[
(1+V_{1}\mu)^{2}-\frac{\gamma_{1}^{2}(1-\mu^{2})}{(1-\mu_{0}^{2})}\right]
^{1/2}\right\}, 
\]
where $E$ and $E_{12}$ are the particle energies (in the region 1 and 
2 rest frames) before and after transmission, respectively, and 
$\gamma_{2}=(1-V_{2}^{2})^{-1/2}$. 
For the transmission of particles from region 2 to region 1,
\[ 
\frac{E_{21}}{E}=\gamma_{1}\gamma_{2}\left\{1+V_{2}\mu-V_{1}\left[
(1+V_{2}\mu)^{2}-\frac{(1-\mu^{2})(1-\mu_{0}^{2})}{\gamma_{2}^{2}}
\right]^{1/2}\right\},
\]
where $E_{21}$ the particle energy (in the region 1 rest frame)  
after transmission.  

The mean acceleration time is defined as $t_{\rm A}=\Delta t/d$, 
where $\Delta t$ is the cycle time and $d$ the mean energy gain 
per interaction. Ignoring particle escape, the cycle time can be 
written in the form  
\begin{equation}
\Delta t=t_{1}P_{\rm r}+(t_{1}+t_{2})P_{12}, 
\end{equation}
where $t_{i}[=2\kappa_{{\rm n}i}/(v_{{\rm n}i}U_{i})]$ denotes the 
mean residence time in region $i$ and $P_{\sigma}$ is the 
probability for process $\sigma$. 
Note that, for $\kappa_{\|}\gg\kappa_{\bot}$, the mean normal velocity  
can be estimated as 
$v_{{\rm n}i}=v_{\|}\sqrt{\kappa_{{\rm n}i}/\kappa_{\| i}}
\simeq v_{\|}\cos\theta_{i}$, where $v_{\|}=v/2\sim c/2$. 

The probabilities of reflection and transmission are expressed as 
$P_{\rm r}=S_{\rm r}/(S_{\rm r}+S_{12})$ and 
$P_{12}=S_{12}/(S_{\rm r}+S_{12})$, respectively. 
Here $S_{\sigma}$ denotes the normal component of particle flux flowing 
into the shock surface, which is calculated by using the normal 
velocity of the guiding centre drift of the particles, that is, 
\[
S_{\sigma}=\int V_{{\rm rel}}d\mu,
\] 
where 
\[
V_{{\rm rel}}=(\mu\cos\theta_{1}+U_{1})/(1-\mu\cos\theta_{1}U_{1})
\]
for $\sigma=$12 and r (in the region 1 rest frame) and
\[ 
V_{{\rm rel}}=(\mu\cos\theta_{2}-U_{2})/(1+\mu\cos\theta_{2}U_{2})
\]
for $\sigma=$21 (in the region 2 rest frame). 
Carrying out the integrations, one approximately obtains 
$P_{\rm r}\simeq\mu_{0}^{2}$ and $P_{12}\simeq 1-\mu_{0}^{2}$. 
Using equations (1) and (2), the cycle time (equation 3) can then be 
expressed as  
\begin{eqnarray}
\Delta t &\simeq&\frac{2\kappa_{{\rm B}}x}{v_{{\rm n}1}U_{1}} \nonumber \\
&\times &\left\{\cos^{2}
\theta_{1}+\frac{\sin^{2}\theta_{1}}{x^{2}}+\frac{r[\cos^{2}\theta_{1}+
\frac{r^{2}}{x^{2}}\sin^{2}\theta_{1}]}{(\cos^{2}\theta_{1}+r^{2}
\sin^{2}\theta_{1})^{3/2}}\right\},
\end{eqnarray}
where $r=U_{1}/U_{2}$ is the shock compression ratio. 
 
The mean energy gain of the upstream particle is denoted as 
\begin{equation}
d=d_{\rm r}P_{\rm r}+(d_{12}+d_{21})P_{12}, 
\end{equation}
where 
\[
d_{\sigma}=\int V_{{\rm rel}}\left(E_{\sigma}/E-1\right)
d\mu/S_{\sigma} 
\]
defines the fractional energy gain in the process 
$\sigma$. To an approximation up to first order in $V_{i}$, 
then, the resultant expressions are 
\[
d_{\rm r}\simeq\frac{4}{3}\mu_{0}V_{1} 
\]
and 
\[
d_{12}=d_{21}\simeq \frac{2}{3}[V_{1}(1-\mu_{0}^{3})/(1-\mu_{0}^{2})-V_{2}]. 
\]
Substituting these results into equation (5) gives 
\[ 
d\simeq\frac{4}{3}[V_{1}-V_{2}(1-\mu_{0}^{2})]. 
\]
As a result, we arrive at the following expression for the mean 
acceleration time (Kobayakawa, Honda \& Samura 2002): 
\begin{eqnarray}
& & t_{\rm A} = \frac{3r\kappa_{\rm B}x}{U_{1}^{2}(r-1)} \nonumber \\
& \times &\left[\cos^{2}\theta_{1}+\frac{\sin^{2}\theta_{1}}{x^{2}}
+\frac{r(\cos^{2}\theta_{1}+\frac{r^{2}}{x^{2}}\sin^{2}\theta_{1})}
{(\cos^{2}\theta_{1}+r^{2}\sin^{2}\theta_{1})^{3/2}}\right]. 
\label{eqn:tobl} 
\end{eqnarray}
Here we have replaced all the downstream quantities with upstream ones. 
Note that equation (6) is valid for the free crossing limit $x\gg 1$. 

In the allowed range of magnetic field inclination angles, 
$\theta_{1}\leq\cos^{-1}U_{1}$ (de Hoffmann \& Teller 1950), 
the value of $t_{\rm A}(\theta_{1}\neq 0^{\circ})$ is smaller than that of 
$t_{\rm A}(\theta_{1}=0^{\circ})$. Relating to the reduction of $t_{\rm A}$, 
we note that $\kappa_{\rm n}$ involved in the acceleration time 
(equation 6) can take a value in the range $> U_{1}f/|\nabla f|$, 
where $f$ is the phase space density of particles. This inequality just 
reflects the condition that the diffusion velocity of particles must 
be larger than the shock speed. Thus, the requirement that the gyroradius 
cannot exceed the characteristic length of the density gradient 
$(f/|\nabla f|)$, recasts the above condition into  
$\kappa_{\rm n}> r_{\rm g}U_{1}$ \citep{jokipii87}. 

\subsection{Particle acceleration without mirror effects}
Following the procedure explained above, we derive the expression 
for the mean acceleration time, excluding mirror effects. 
Assuming that all upstream particles are transmitted downstream 
through the shock front (no reflection), thereby, setting  
$P_{\rm r}=0$ and $P_{12}=1$ in equation (3), reduces the 
expression of the cycle time to 
\begin{equation}
\Delta t^{\prime}=t_{1}+t_{2}. 
\end{equation}
Equation (7) can be explicitly written as 
\begin{eqnarray}
& &\Delta t^{\prime}\simeq\frac{2\kappa_{{\rm B}}x}{v_{n1}U_{1}} \nonumber \\
&\times&\left\{\cos^{2}\theta_{1}
+\frac{\sin^{2}\theta_{1}}{x^{2}}+\frac{r[\cos^{2}\theta_{1}+
(r^{2}/x^{2})\sin^{2}\theta_{1}]}{\cos^{2}\theta_{1}+r^{2}
\sin^{2}\theta_{1}}\right\},
\end{eqnarray}
for $x\gg 1$. 
Note that the difference from equation (4) is only the denominator of 
the third term in the curly brackets. 

Similarly, the mean energy gain per interaction is denoted as    
\begin{equation}
d^{\prime}=d_{12}+d_{21}. 
\end{equation}
Recalling that all particles having pitch angle of $0<\mu\leq 1$ 
are forced to be transmitted to region 2, we have 
\[
d^{\prime}\simeq \frac{4}{3}(U_{1}-U_{2}). 
\]
In contrast to $d(\propto1/\cos\theta_{1})$, the expression for 
$d^{\prime}$ is independent of the field inclination, 
and appears to be identical to that for parallel shocks. 
Using equations (8) and (9), the acceleration time without mirror effects, 
defined by $t_{\rm A}^{\prime}=\Delta t^{\prime}/d^{\prime}$, is found to 
be represented in the following form: 
\begin{eqnarray}
& & t_{\rm A}^{\prime} =\frac{3r\kappa_{\rm B}x}{U_{1}^{2}(r-1)
\cos\theta_{1}} \nonumber \\
&\times&\left\{\cos^{2}\theta_{1}+\frac{\sin^{2}\theta_{1}}{x^{2}}
+\frac{r[\cos^{2}\theta_{1}+(r^{2}/x^{2})\sin^{2}\theta_{1}]}
{\cos^{2}\theta_{1}+r^{2}\sin^{2}\theta_{1}}\right\}.  
\label{eqn:twom}
\end{eqnarray}

In comparison with equation (6), the value of equation (10) is boosted 
as a result of the factor $(\cos\theta_{1})^{-1}$ and the smaller 
denominator of the third term in the curly brackets. The latter 
comes directly from exclusion of mirror effects. The factor 
$(\cos\theta_{1})^{-1}$ reflects the anisotropy of particle velocity, 
involved in the mean residence time $t_{i}$. Although in 
the evaluation of $t_{\rm A}$ this factor was canceled out by the 
same factor from the expression for $d$, in the present case it is 
not cancelled, because of the independence of $\theta_{1}$ in 
$d^{\prime}$. Note the relation
\[ 
t_{\rm A}^{\prime}(\theta_{1}=0^{\circ})=t_{\rm A}(\theta_{1}
=0^{\circ})=(3\kappa_{\rm B}x/U_{1}^{2})[r(r+1)/(r-1)], 
\]
which coincides with the acceleration time for a parallel shock. 

\section{Numerical results}
\begin{figure}
\includegraphics[width=90mm]{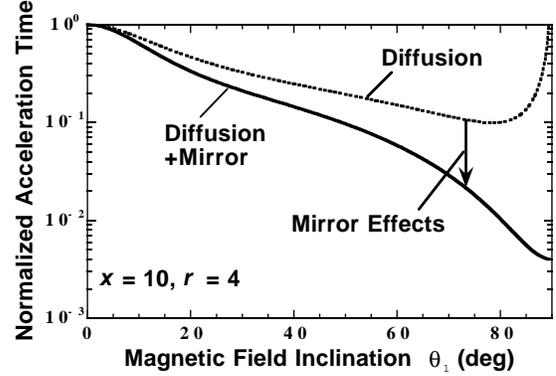}
\caption{The acceleration times for an oblique shock normalized by 
that for a parallel shock, as a function of the magnetic field inclination 
$\theta_{1}$ (degrees) with respect to the shock normal direction. Here, 
a strong shock ($r=4$) and weak turbulence ($x=10$) have been assumed. 
The normalized time including the effects of both mirror 
reflection and diffusion anisotropy $\tilde{t_{\rm A}}$ is represented 
by a solid curve and that including solely the effects of diffusion 
anisotropy $\tilde{t_{\rm A}}^{\prime}$ by a dotted one. The 
reduction of the time-scale indicated by the arrow is ascribed to 
mirror effects, which are pronounced in the large-$\theta_{1}$ region.}
\label{fig:f1}
\end{figure}

Below, we assume a monatomic, non-relativistic gas with specific heat 
ratio of $5/3$, whereby $1<r\leq 4$ for a non-relativistic shock. In 
Fig.~\ref{fig:f1}, we display the $\theta_{1}$ dependence of equations 
(6) and (10), in the case of $r=4$ for the strong shock limit and 
$x=10$ compatible with the assumption of weak turbulence ($x\gg 1$). 
The upper dotted curve denotes the acceleration time without mirror 
effects normalized by that for a parallel shock: 
\[
\tilde{t_{\rm A}}^{\prime}=t_{\rm A}^{\prime}(\theta_{1})/
t_{\rm A}^{\prime}(\theta_{1}=0^{\circ}). 
\]
The favoured reduction of the acceleration time for 
$\theta_{1}\neq 0^{\circ}$ stems from the effects of diffusion 
anisotropy. It is noted that, for $\theta_{1}\neq 0^{\circ}$, the reduction 
of the shock normal component of the particle velocity, $v_{{\rm n}i}$, 
increases the mean residence time $t_{i}$, and thus degrades the 
shock-crossing. As seen in Fig.1, this effect dominates the 
anisotropic diffusion effect (coming from the smaller 
$\kappa_{{\rm n}i}$) for $\theta_{1}\ga 80^{\circ}$,
where $\tilde{t_{\rm A}}^{\prime}$ changes to an increase. In the limit of 
$\theta_{1}\rightarrow 90^{\circ}$, $\tilde{t_{\rm A}}^{\prime}$ 
diverges, because of the related factor $(\cos\theta_{1})^{-1}$ 
in equation (10) (see Section 2.3). Hownever, note that the present 
calculations are physically meaningful only for inclination 
$\theta_{1}\leq \cos^{-1}(U_{1})$, as mentioned in Section 2.2. 
For comparison, the normalized acceleration time including the effects 
of both the mirror reflection and diffusion anisotropy, 
\[  
\tilde{t_{\rm A}}=t_{\rm A}(\theta_{1})/t_{{\rm A}}(\theta_{1}=0^{\circ}),
\]  
is also plotted (solid curve). 
That is, the further reduction of the acceleration time (from dotted to 
solid) for $\theta_{1}\neq 0^{\circ}$ can be ascribed to the effects of 
mirror reflection. It is found that the contribution of diffusion 
anisotropy is larger for smaller $\theta_{1}$, whereas, for 
larger $\theta_{1}$, mirror effects play a dominant role in reducing 
the acceleration time. 
At $\theta_{1}=90^{\circ}$, instead of the mirroring, the shock 
drift mechanism actually dominates (e.g., Webb, Axford \& Terasawa 1983), 
violating the present formalism. 

\begin{figure}
\includegraphics[width=90mm]{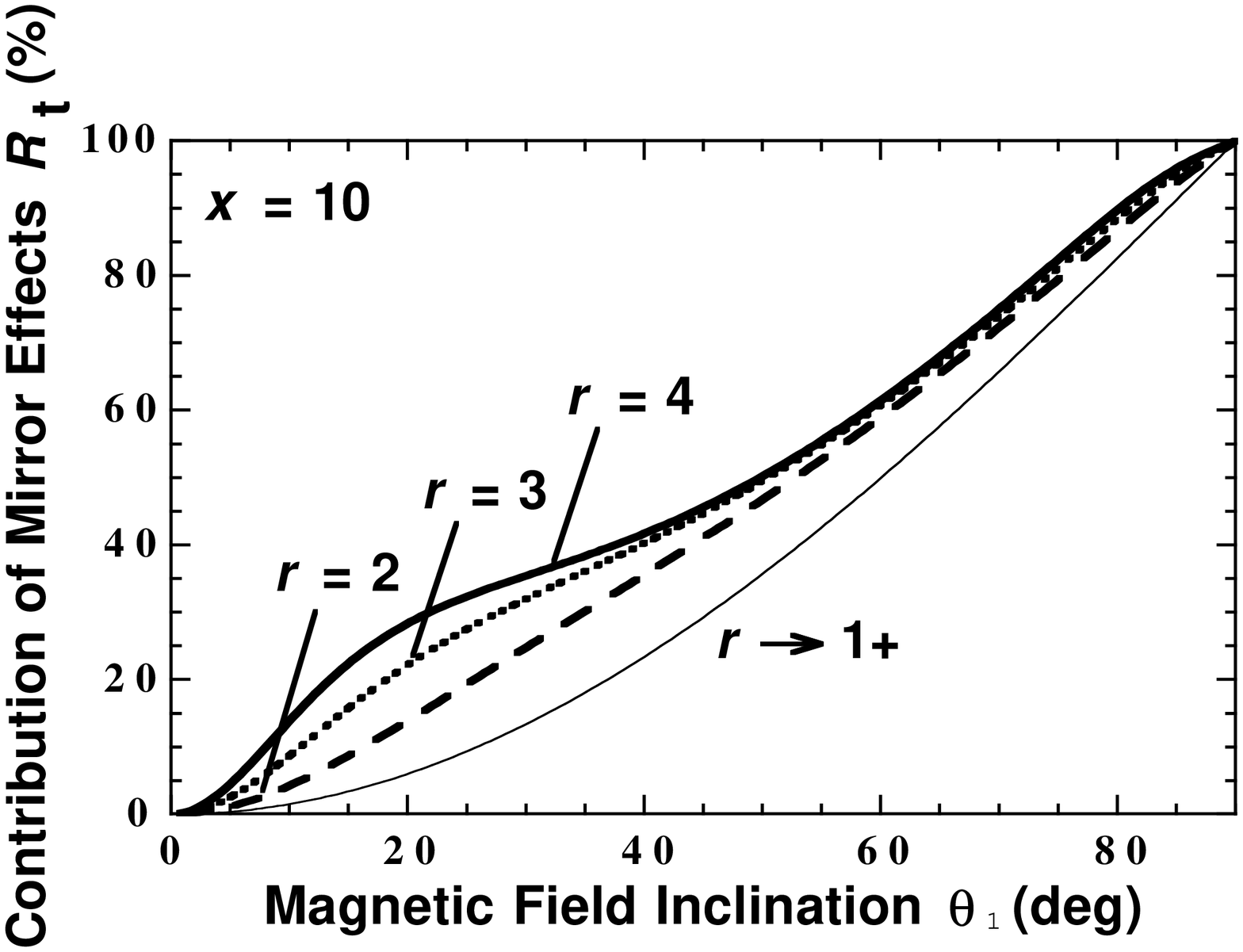}
\caption{The contribution rate of mirror effects $R_{\rm t}$ 
(per cent) against the magnetic field inclination $\theta_{1}$ (degrees) 
for $x=10$. 
The cases with the shock compression ratio of $r\rightarrow 1+$ 
and $r=2$, $3$ and $4$ are denoted by thin solid, dashed, dotted  
and thick solid curves, respectively.}
\label{fig:f2}
\end{figure}

In order to give a clear presentation of the results, we define the 
contribution rate of the mirror effects as
\[ 
R_{\rm t}=|t_{\rm A}^{\prime}-t_{\rm A}|/t_{\rm A}^{\prime}\times 100
~~~({\rm per~ cent}). 
\]
Note that this rate is dependent on $\theta_{1}$, $x$, and $r$,  
and independent of the other parameters. 
In Fig.~ \ref{fig:f2} for $x=10$, we plot $R_{\rm t}$ as a function of 
$\theta_{1}$, given $r$ as a parameter in the range $1<r\leq 4$. 
We mention that, for $x\geq 10$, the values of $R_{\rm t}$ do not 
change very much over the whole range of $\theta_{1}$. For example, 
in the case of $r=4$, the difference in the $R_{\rm t}$ values  
between $x=10$ and $100$ is at most 4.3 per cent at $\theta_{1}=79^{\circ}$ 
(not shown in the Figure). 
In the special case of $\theta_{1}=0^{\circ}$ (parallel shock), 
the effects of both mirror reflection and anisotropic diffusion vanish, 
so that $R_{\rm t}=0$ per cent irrespective of the compression ratio. 
As $\theta_{1}$ increases, the values of $R_{\rm t}$ increase monotonically,  
and reach nearly 100 per cent at $\theta_{1}\sim 90^{\circ}$
(quasi-perpendicular shock). 
As would be expected, 
the contribution of mirror effects is larger in a stronger shock. 
For the $r=4$ case, the $R_{\rm t}$ value reaches 50 per cent 
at $\theta_{1}\simeq 50^{\circ}$ and 
exceeds 80 per cent at $\theta_{1}\simeq 74^{\circ}$.  
On the other hand, for $r\rightarrow 1+$ (weak shock limit), 
$R_{\rm t}=50$ per cent can be achieved at $\theta_{1}=60^{\circ}$. 
The difference in $R_{\rm t}$ for the strong shock case 
from that for the weak shock case is more pronounced in relatively 
low inclination angles. It is also found that, in the range  
$\theta_{1}\sim 50^{\circ}-70^{\circ}$, the $R_{\rm t}$ values 
merely vary slightly for $r\geq 2$.  
Anyway, we can claim that the mirror reflection is effective 
in quasi-perpendicular shocks.

\section{Conclusions}
For a non-relativistic, fast-mode oblique shock accompanied by  
weak MHD turbulence, we have quantitatively revealed the contribution 
of magnetic mirror effects and anisotropic diffusion effects to 
the reduction of the acceleration time of cosmic ray particles. 
We found in particular that, in the strong shock limit, for a magnetic 
field inclination angle (to the shock normal) of 
$\theta_{1}>50^{\circ}$, mirror effects contribute dominantly 
to the reduction of the acceleration time; whereas, for 
$\theta_{1}<50^{\circ}$, anisotropic diffusion effects contribute 
dominantly to that time. In the small-$\theta_{1}$ region, the contribution 
rate of mirror effects is found to be small, but sensitive to the shock 
strength, such that a larger shock compression leads to a more enhanced 
contribution rate. 

While these consequences can be directly referred to the study of 
oblique shock acceleration in space and heliospheric environments, 
one should be a little more careful for an application to 
other objects, including supernova remnants. We also remark 
that the perpendicular diffusion of cosmic rays is still 
not well understood in many astrophysical aspects. In a common approach, 
the diffusion coefficient is related to the spectral intensity of 
magnetic field fluctuations. For example, spectral analysis of 
fluctuations in the solar corona shows that,  
in the region of the heliocentric radius of $3R_{\odot}<R<6R_{\odot}$, 
the power-law indices can be fitted by $\nu\simeq 1.6$, which can be 
approximated by $5/3$ for the Kolmogorov turbulence,  
and in the $6R_{\odot}<R<12R_{\odot}$ region, $\nu\simeq 1.1$,  
which can be approximated by $1$ for the Bohm diffusion limit 
\citep{chashei00}. In interplanetary 
gas clouds (extrasolar planetary systems), the power 
spectrum has also been characterized in analogy with 
the description of Kolmogorov turbulence \citep{watson01,wiebe01}. 
Although fluctuations have been confirmed in such various objects, 
this does not always mean that the estimated values 
of $x$ are pertinent to the present scheme. 

In young shell-type supernova remnants, especially, strong amplification 
of downstream magnetic field has been confirmed by recent observations 
(e.g. Bamba et 
al. 2003, Vink \& Laming 2003). Filamentary structures of hard X-rays are 
interpreted as evidence of shock modification predicted by the 
non-linear theory (e.g. Berezhko, Ksenofontov \& V\"olk 2003, Berezhko \& 
V\"olk 2004, V\"olk, Berezhko \& Ksenofontov 2005), and analytical studies 
and numerical simulations of plasma instability \citep{bell04}. The 
expected $x$ values of these sources are of the order of unity, which arguably 
leads to effective acceleration, though the present formalism 
breaks down in the diffusive limit of $x=1$, where the magnetic field 
inclination fluctuates largely, to be ambiguously defined. Moreover, 
it is not appropriate to use the unperturbed trajectory of the guiding 
centre drift motion as an approximation while particles are reflected 
at the shock surface by magnetic mirroring. The relevant issues, 
departing from the free crossing approximation, are beyond the scope of this 
paper. 

\bsp

\label{lastpage}

\end{document}